\documentclass[aps,english,prl,amsmath,amsfonts,amssymb,superscriptaddress,twocolumn,citeautoscript]{revtex4}

\usepackage[version=3]{mhchem} 
\usepackage[T1]{fontenc}
\usepackage[latin9]{inputenc}
\usepackage{amsmath}
\usepackage{amssymb}
\usepackage{wasysym}
\usepackage{esint}
\usepackage{graphicx}
\usepackage{times}
\usepackage{nicefrac}
\usepackage{appendix}
\usepackage{textcase}
\usepackage{subfigure}
\usepackage{empheq}
\usepackage{color}
\usepackage{leftidx}
\usepackage{makecell}
\usepackage{stackrel}
\usepackage{soul}
\makeatletter

\renewcommand{\b}{\beta}

\newcommand{\add}[1]{\if\a\b{{\color{red} #1}}\else{#1}\fi}

\renewcommand{\eqref}[1]{Eq.~\ref{eq:#1}}

\newcommand{\figref}[1]{Fig.~\ref{fig:#1}}

\makeatother

\usepackage{babel}

\begin{document}
\title{Laser cooling assisted thermal management of lightsails}
\author{Weiliang Jin}
\affiliation{Department of Electrical Engineering, Ginzton Laboratory, Stanford University, Stanford, California 94305, USA}
\author{Wei Li}
\affiliation{Department of Electrical Engineering, Ginzton Laboratory, Stanford University, Stanford, California 94305, USA}
\affiliation{GPL Photonics Laboratory, State Key Laboratory of Applied Optics, Changchun Institute of Optics, Fine Mechanics and Physics, Chinese Academy of Sciences, Changchun 130033, China}
\author{Chinmay Khandekar}
\affiliation{Department of Electrical Engineering, Ginzton Laboratory, Stanford University, Stanford, California 94305, USA}
\author{Meir Orenstein}
\affiliation{Department of Electrical Engineering, Technion-Israel Institute of Technology, 32000 Haifa, Israel}
\author{Shanhui Fan}
\affiliation{Department of Electrical Engineering, Ginzton Laboratory, Stanford University, Stanford, California 94305, USA}

\begin{abstract}
  A lightsail can be accelerated to relativistic speed by the
  radiation pressure of a laser having an intensity of the order of
  GW/m$^2$. Such an extreme light intensity presents a critical
  challenge in the thermal management of lightsails. In this letter,
  we propose to use solid-state laser cooling for dissipating heat
  from such a lightsail. Our approach uses the same laser that is
  accelerating the sail, and can be used in addition to the previously
  explored radiative cooling. With our approach, we show that the
  cooling rate of a sail composed of a micron-thick layer doped with
  ytterbium ions can exceed that of blackbody thermal emission. This
  allows illumination by higher intensity lasers, and consequently
  shortens the acceleration distance to reach relativistic speed. Due
  to the Doppler shift, the performance of our approach is impacted by
  the limited bandwidth of the laser cooling dopant. When a
  constant-frequency pumping laser is used, laser cooling is
  particularly helpful for target velocities $\lesssim0.05c$ for
  room-temperature operations.
\end{abstract}
\maketitle

Recent advances in high-power laser technology have stimulated
progress in both fundamental physics and many applications relying on
strong laser-matter interactions~\cite{salamin2006relativistic}. For
instance, intense laser pulses have been explored to accelerate
electrons~\cite{malka1997experimental,wootton2016demonstration} and
ions~\cite{esirkepov2004highly} to a relativistic speed. More
recently, there is a growing interest in accelerating macroscopic
objects to high speeds as well. Notable examples include fast-transit
earth orbital maneuvering and interplanetary
flight~\cite{tung2021light}, and the Starshot Breakthrough Initiative
that aims to accelerate a meter-size lightsail to 20$\%$ of the speed
of light, so that it can reach a nearby star Proxima Centauri in 20
years~\cite{lubin2016roadmap, parkin2018breakthrough}. Many aspects of
lightsail design have been
discussed~\cite{salary2020photonic,atwater2018materials}, including
efficient
propulsion~\cite{atwater2018materials,Jin2020,kudyshev2021optimizing},
thermal management~\cite{ilic2018nanophotonic}, and self-stabilization~\cite{ilic2019self, siegel2019self, manchester2017stability,
  myilswamy2020photonic}.

A critical issue related to the necessary use of high-power laser for
lightsail acceleration is the heat generated due to the residual
optical absorption, even for a very low-loss dielectric sail, and the
resultant heat that has to be dissipated efficiently to avoid material
damage~\cite{liu2006thermal, bowman1999lasers}. This is exasperated by
the fact that material absorption in general increases with
temperature, causing significant thermal
runaway~\cite{holdman2021thermal}. The challenge is becoming more
critical in space where heat cannot be dissipated through conduction
or convection. Radiative cooling, which is based on thermal radiation,
is thus the only approach explored for cooling a
lightsail~\cite{salary2020photonic, ilic2018nanophotonic}. In this
context, simultaneous optimization for laser propulsion and thermal
emission have been performed~\cite{brewer2021multi}, but implementing
the resulting photonic design over a meter-scale surface for
emissivity enhancement remains challenging.

On the other hand, the same laser beam that causes heating might be
exploited to provide cooling by applying the concept of solid-state
laser cooling based on anti-Stokes
fluorescence~\cite{seletskiy2012cryogenic, nemova2015twenty,
  seletskiy2016laser}. Laser cooling in solids has been successfully
demonstrated in several materials including rare-earth ion doped
fluoride glass, crystals~\cite{seletskiy2016laser,
  seletskiy2012cryogenic} and silica glass~\cite{knall2020laser,
  mobini2020laser, peysokhan2020laser}, and is extensively pursued in
semiconductors~\cite{huang2004theoretical, zhang2019progress,
  sheik2009laser}. One application of laser cooling is to provide
solid-state refrigeration from room temperature to cryogenic
temperatures, requiring the rare-earth ions to have near-unity quantum
yield and the host media to be crystalline to enhance the absorption
cross section of the ions~\cite{nemova2015twenty}. A remarkable low
temperature of $91$~K has been experimentally
achieved~\cite{melgaard2016solid}. Another application is related to
high-power lasers. The concept of radiation-balanced fiber
lasers~\cite{bowman1999lasers, knall2018model, balliu2019predictive,
  knall2020laser} has been proposed to balance the heat generated by
the Stokes-shifted lasing process with anti-Stokes fluorescence. The
light intensity inside the fiber can be as high as $\sim$GW/m$^2$,
approaching the value expected in the lightsail setups. Laser cooling
operating at elevated temperature has also been
discussed~\cite{nakayama2019improving}, where an improved cooling
efficiency is to be expected.

In this article, we explore the possibility of adopting solid-state
laser cooling based on rare-earth ions to address the thermal
management challenge in the lightsail, which will potentially allow
for more intense laser power and therefore shorter acceleration
distance for the lightsail to reach the target velocity. Compared to
prior laser cooling applications, there are at least three unique
aspects in the lightsail platform: the operating temperature being
higher than previously explored values; the laser intensity flux
substantially saturating the absorption of the rare-earth ions; and
the laser frequency experiencing significant Doppler redshift in the
lightsail frame during the acceleration. The choice of the materials
and the design of structures need to take these unique aspects into
account. Here we develop a formalism for the design of laser-cooled
lightsails. This formalism characterizes the performance of the laser
cooling by the thermally constrained acceleration distance, with which
the dual interplay between the optical and the thermal processes are
captured. The minimization of acceleration distance is crucial for
lowering both the power consumption and laser array
size~\cite{ilic2018nanophotonic}. With our formalism, we show that
laser cooling in micron-thick structures can offer substantial cooling
power compared to blackbody emission, which leads to large improvement
for the acceleration distance (potentially, up to $\gtrsim 20 \times$
in the ideal regime, or $\gtrsim 10\times$ for the state-of-art). We
explore the dependence on various parameters, including the material's
absorption and the lightsail's temperature and target velocity. We
find that laser cooling, in the simplest settings, is particularly
relevant for lower temperatures and target velocities of few percents
the speed of light. We finally study the possibility of having a
variable laser intensity during the acceleration stage, which makes
the most benefit of the laser cooling power and gives additional
improvement for the acceleration distance.

\section{Theory}

\begin{figure}[htbp]
  \centering
  \includegraphics[width=0.9\linewidth]{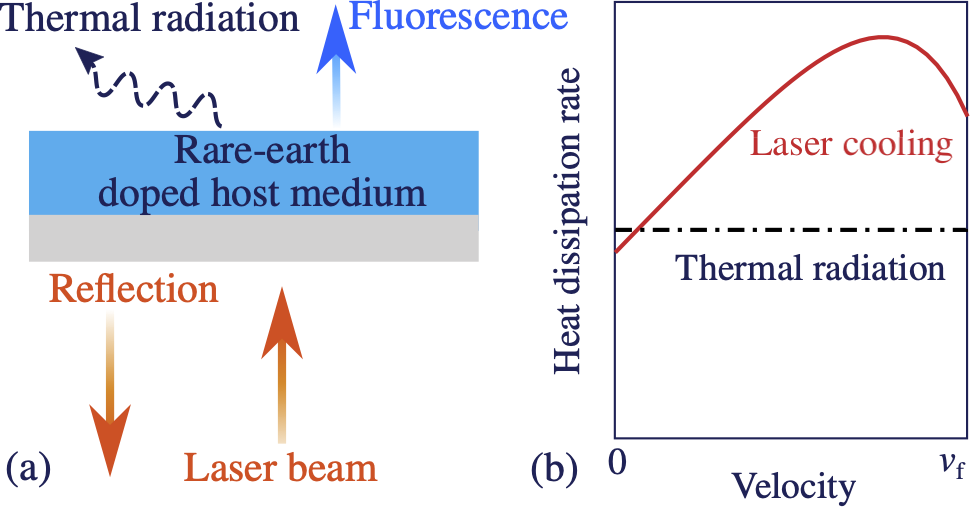}
  \caption{(a) Schematic of the concept of a laser-cooled
    lightsail. Most energy of the incident laser beam is reflected and
    contributes to the optical force for the propulsion, while some
    light power is parasitically absorbed and leads to heating. The
    heat is dissipated by thermal radiation, as well as via the
    anti-Stokes emission from the laser-excited rare-earth ions with
    fluorescence frequency higher than the incident laser
    frequency. (b) Schematic of total heat dissipation rate from the
    lightsail held at a constant temperature as a function of the
    velocity of the lightsail. All quantities are defined in the
    lightsail frame. The rate is constant for thermal radiation alone
    (black dotted-dashed curve), but exhibits variations with laser
    cooling (red solid curve) due to relativistic effects on the
    received laser intensity and Doppler-shifted
    frequency~\cite{ilic2018nanophotonic} that can alter laser cooling
    rate described by \eqref{cool} in the main text.}
  \label{fig:scheme}
\end{figure}

As schematically illustrated in \figref{scheme}, a high-power laser
beam is incident on a multi-layered lightsail, with at least one layer
doped with rare-earth ions. The laser beam provides propulsion through
radiation pressure, but nevertheless heats up the lightsail by its
residual optical absorption. To balance against the heat generation, a
portion of the laser light is used to pump the rare earth ions, which
results in heat removal by anti-Stokes fluorescence on the top of the
native thermal radiation. Below we briefly discuss formulations of the
aforementioned processes, which are required to the study the
acceleration and thermal balance of the lightsail.

The lightsail propulsion, as established in previous
research~\cite{chen2011optical,ilic2018nanophotonic}, depends
crucially on the reflectivity that determines the momentum transfer,
and the mass density that impacts the acceleration rate. A
comprehensive figure of merit that captures the trade-off between the
two is the distance $D$ for the lightsail to be accelerated to a
target velocity, described by the following
equation~\cite{atwater2018materials,kulkarni2018relativistic,
  Jin2020},
\begin{equation}
D = \frac{c^3}{2}(\rho_l+\rho_s)\int_0^{\beta_f} \frac{1}{I_p(\beta)}  \frac{\gamma\beta\mathrm{d}\beta}{R[\lambda_{\beta}](1-\beta)^2} \label{eq:D}
\end{equation}
In \eqref{D}, $c$ is the speed of light, $\beta=v/c$ the velocity
normalized by $c$, $\gamma=1/\sqrt{1-\beta^2}$ the Lorentz factor,
$\beta_f$ the target velocity, $\rho_{s}$ ($\rho_{l}$) the area mass
density of the lightsail (payload), $I_p$ the laser intensity averaged
over the lightsail area when observed in the earth frame (where we
explicitly included the $\beta$ dependence for when the intensity is
made variable during the acceleration stage). The reflectivity $R$,
defined in the lightsail frame, is a function of the laser wavelength
$\lambda$ that experiences Doppler redshift from the laser wavelength
in the rest frame $\lambda_p$ as described by
$\lambda_{\beta}=\lambda_p\sqrt{(1+\beta)/(1-\beta)}$.

\eqref{D} suggests that $D$ can be decreased by simply boosting the
laser intensity $I_p$. However, the optical power absorption of the
lightsail also increases with $I_p$, which due to limited heat
dissipation channels, can significantly raise the temperature of the
lightsail. To avoid material damage, the highest temperature during
the entire propulsion
process~\cite{atwater2018materials,ilic2018nanophotonic} must be below
a threshold temperature $T_s$. Such a thermal management consideration
imposes an upper limit on the allowed laser intensity, denoted
$I_{p,\mathrm{max}}$. Other sources of material damage under high-power laser illumination such as plasma
formation and ablation are typically reserved for
short-pulses~\cite{stuart1995laser} and are not considered here. $I_{p,\mathrm{max}}$ can be found
by considering the thermal balance in the co-moving frame of the
lightsail,
\begin{equation}
  I_{\mathrm{abs}}(I_\textrm{sail},\beta)\leq I_{\mathrm{th}}(T_s)+I_{\mathrm{cool}}(I_\textrm{sail},T_s,\beta),\; \; \forall \beta\in[0,\beta_f],\label{eq:balance}
\end{equation}
throughout the entire acceleration stage, where
$I_{\textrm{sail}}(\beta) = I_p\frac{1-\beta}{1+\beta}$ is the laser
intensity in the lightsail frame~\cite{ilic2018nanophotonic}, and
$I_{\mathrm{abs}}$, $I_{\mathrm{th}}$, and $I_{\mathrm{ion}}$ denote
the power flows per unit area corresponding to optical absorption,
thermal radiation, and laser cooling respectively. More explicitly,
the absorbed power per unit area can be expressed as
$I_{\mathrm{abs}}(\beta) = I_{\textrm{sail}}(\beta)
A(\lambda_{\beta})$ that directly depends on
$I_{\textrm{sail}}(\beta)$, and $A(\lambda_{\beta})$, the absorptivity
of the lightsail.

The computation of thermal radiation from a multilayer structure is
well established, which can be expressed as~\cite{ilic2018nanophotonic}
\begin{align}
  &I_{\mathrm{th}}(T_s) = \int_0^{\infty}\mathrm{d}\lambda \varepsilon(\lambda)\pi I_{\mathrm{BB}}(\lambda,T_s)\label{eq:thermal}
\end{align}
where $\varepsilon(\lambda)$ is the hemispherical-spectral emissivity
summing over the front and back surface, and
$I_{\mathrm{BB}}(\lambda,T_s)=\frac{2hc^2}{\lambda^5}\left[
  \exp(\frac{hc}{\lambda k_BT_s})-1 \right]^{-1}$ the blackbody
spectral intensity.

Finally, we compute $I_{\mathrm{cool}}(I_\textrm{sail},T_s,\beta)$
using the recently developed adaptive four-level
model~\cite{jin2021adaptive} that captures the main features of laser
cooling operating at elevated temperatures. This simple
phenomenological model, validated against experimental results, can
accurately predict the cooling performance over a wide range of
temperatures and pumping frequencies using only a small number of
parameters that can be determined from experiments. This is
particularly relevant to the lightsail problem since the pumping
wavelength can undergo significant Doppler shift
[\figref{scheme}(b)]. The rare-earth ions, responsible for the laser
cooling, absorb light over a narrow frequency range since the
bandwidths of the excited and ground states manifold that contribute
to the absorption are relatively narrow. The laser cooling rate is
\begin{align}
  &P_{\mathrm{cool}}(I,T_s,\omega_p) = \left[ \eta_e \frac{\omega_f}{\omega_p}-1 \right]\alpha(I,T_s,\omega_p)I\label{eq:cool}
\end{align}
where $\eta_e$ is the external quantum yield and
$\alpha(I,T_s,\omega_p)$ is the absorption coefficient of the ions
that depends on the pumping frequency $\omega_p$ and the local light
intensity $I$ at the position of ions, as follows,
\begin{align}
  &\alpha(I) = \frac{\alpha_0}{1+I/I_s}, \;\mathrm{with}\;\;  \alpha_0 = \frac{\sigma_{12}N_t}{1+\exp[\hbar\Delta/k_BT_s]}.\label{eq:alpha}
\end{align}
$\sigma_{12}$ is the atomic absorption cross section for the
transition from the top of the ground manifold to the bottom of the
excited manifold, $N_t$ the doping concentration,
$I_s=\hbar\omega\gamma/\sigma_{12}$ the saturation intensity. $\Delta$
is the effective manifold width, which is a solution of the following
equation, assuming a given mean fluorescence frequency $\omega_f$,
\begin{equation}
  \omega_f = \omega_{p} + \Delta \left[ \frac{1}{2}+\frac{1}{1+\exp[\hbar\Delta/k_BT_s]} \right].\label{eq:mean}
\end{equation}
Neglecting fluorescence reabsorption~\cite{heeg2005influence}, the net
cooling power $I_{\mathrm{cool}}$ needed in \eqref{balance} can then
be computed simply by integrating $P_{\mathrm{cool}}$ over the layer
thickness $t$, which gives an increasing $I_{\textrm{cool}}$ with
thickness. This is the relevant regime for thin lightsails where only
a small fraction of the incident light is absorbed. Finally, note that
in laser cooling setups, there is an additional parasitic absorption
term $\alpha_bI$ which is not included in \eqref{cool}, but is
directly accounted for in the optical absorption $I_{\textrm{abs}}$ in
\eqref{balance}. This parasitic absorption contributes to heating,
while the absorption $\alpha$ in \eqref{cool} leads to cooling.

We note that \eqref{cool} can be simplified in the high intensity
regime, which is relevant for lightsail platforms where the typical
laser intensity ($I_p\gtrsim10$~GW/m$^2$~\cite{lubin2016roadmap,
  atwater2018materials}) is far above the saturation intensity
($I_s\approx0.1$~GW/m$^2$~\cite{knall2020laser} for typical parameters
of rare-earth ions, including
$\gamma\approx10^3$~s$^{-1}$~\cite{digonnet2001rare} and
$\sigma_{12}\approx 10^{-20}$~cm$^{-2}$~\cite{knall2020laser}). In
this limit, the cooling power density yields a simpler expression,
\begin{align}
  P_{\mathrm{cool}}=\left( \eta_e \frac{\omega_f}{\omega_p}-1 \right) \frac{N_t\hbar\omega_p\gamma}{1+\exp[\hbar\Delta/k_BT_s]}\label{eq:sat}
\end{align}
Therefore, $P_{\mathrm{cool}}$ depends linearly on the decay rate
$\gamma$ and the doping concentration $N_t$, but is independent on the
light intensity $I$. $P_{\mathrm{cool}}$ is also independent of
$\sigma_{12}$, which is very different from laser cooling applications
pursuing cryogenic refrigeration whose performances can be greatly
improved by enhancing $\sigma_{12}$~\cite{seletskiy2012cryogenic}. In
the ideal scenarios where $\eta_e\approx1$, the peak cooling power
density can be found by solving for $\omega_p$ that maximizes
\eqref{sat},
\begin{align}
  &P_{\mathrm{cool,m}}\approx 0.21\gamma k_BT_s N_{t}\nonumber\\
  &I_{\mathrm{cool,m}}  \approx 0.21\gamma k_BT_s N_{t}t \mathrm{,\; at}\;\omega_p\approx\omega_f-0.78 \frac{k_BT_s}{\hbar},\label{eq:peak}
\end{align}
where the corresponding relative bandwidth of the cooling spectrum,
characterized by normalized full width at half maximum, is
$1.24 \frac{k_BT_s}{\hbar\omega_p}$. This finite bandwidth can limit
the laser cooling performance of a lightsail, as we explore later
[\figref{scheme}(b)].

\section{Results}

\begin{figure}[htbp]
  \centering
  \includegraphics[width=0.44\textwidth]{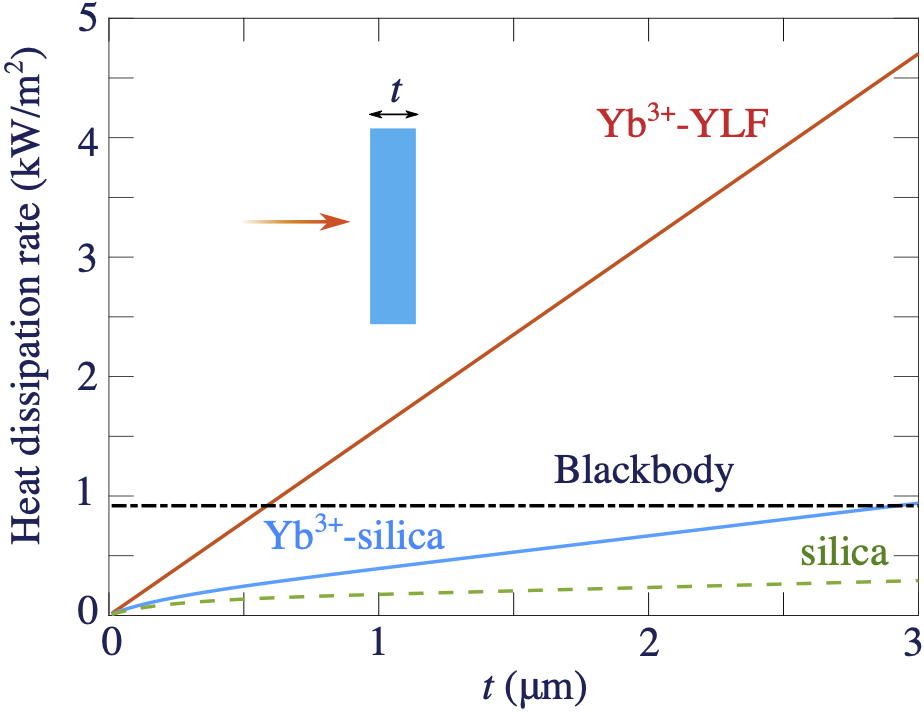}
  \caption{Heat dissipation rate per unit area at room temperature for
    ideal Yb$^{3+}$-doped YLF (red solid line) and Yb$^{3+}$-doped
    silica (blue solid line) in the saturated absorption regime where
    the laser intensity is far above the saturation intensity
    throughout the slab, compared to the heat dissipation rate due to
    blackbody emission (black dashed line) and thermal emission alone
    from silica slab (green dashed line). Note that the blue solid
    curve is for the summed contribution from thermal radiation and
    laser cooling. The doping level is assumed to be
    $1.4\times10^{21}$~cm$^{-3}$ for YLF ~\cite{melgaard2016solid} and
    $1.93\times10^{20}$~cm$^{-3}$ for
    silica~\cite{knall2021radiationPRL}, and the radiative lifetime is
    $\tau=0.765$~ms~\cite{knall2021radiationPRL}.}
  \label{fig:power}
\end{figure}

In order to study the relevance of laser cooling in a lightsail, we
first examine a simple single-slab geometry and compute heat
dissipation rates due to thermal radiation and laser cooling for
different materials (\figref{power}). We find that in the saturated
regime (\eqref{peak}), laser cooling (solid lines) can substantially
improve the heat dissipation rate compared to thermal radiation alone
(dahsed lines). In particular, due to laser cooling, a Yb$^{3+}$-doped
silica slab (blue line) has a larger heat dissipation rate compared to
an undoped layer that only relies on thermal radiation (dashed
line). The heat dissipation rate is more than doubled for a
$3~\mu$m-thick layer assuming state-of-art doping level (indicated in
the caption). However, even with the increase due to laser cooling,
the total heat dissipation rate may still be considered relatively
moderate for the lightsail application. On the other hand, a
significantly larger heat dissipation rate is observed for a
Yb$^{3+}$-doped $\textrm{LiYF}_4$ (YLF) layer (red line) in the
optimal laser cooling regime with a state-of-art doping level. With
this material, the total dissipation rate can significantly exceed
blackbody emission (dot-dashed line) using a micron-thick layer. This
promising result suggests a large potential of improved thermal
management in lightsails using laser cooling.

\begin{figure}[h]
  \centering
  \includegraphics[width=0.44\textwidth]{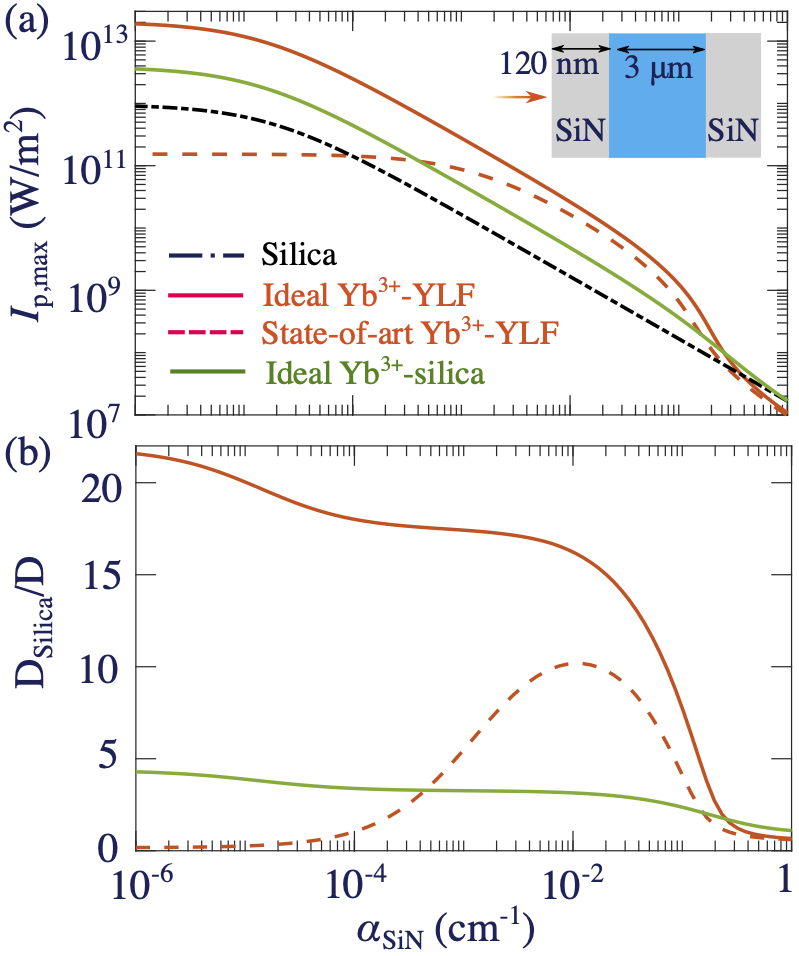}
  \caption{The maximum allowed pumping intensity (a) and improvement
    of the acceleration distance (b) using laser cooling (where the
    middle layer is either Yb$^{3+}$-doped silica [green line] or
    Yb$^{3+}$-doped YLF crystal [red curve]) compared to using undoped
    silica layer as a middle layer. We assume constant laser intensity
    during the acceleration stage. ``Ideal'' laser-cooled layer
    corresponds to $\eta_e=100\%$ and
    $\alpha_b=\alpha_{\textrm{silica}}=10^{-6}~\textrm{cm}^{-1}$ (we
    take a non-zero background absorption for fair comparison with
    undoped silica). We use the state-of-art parameters for
    Yb$^{3+}$-YLF ($\eta_e=99.6\%$, $\alpha_b=10^{-4}$~cm$^{-1}$,
    radiative lifetime $\tau=0.765$~ms,
    $\sigma_{12}=2.8\times10^{-21}$~cm$^{-2}$) and only consider the
    ideal case for Yb$^{3+}$-silica (state-of-art gives relatively
    poor performance). The doping level is assumed to be the
    state-of-art in silica
    $1.93\times10^{20}$~cm$^{-3}$~\cite{knall2021radiationPRL} and YLF
    $1.4\times10^{21}$~cm$^{-3}$~\cite{melgaard2016solid}. Thermal
    radiation from both SiN and silica layers are considered.}
  \label{fig:power2}
\end{figure}

In order to study more carefully the advantages of a lightsail
application, we compute the resulting effects on the acceleration
distance (\eqref{D}). While both doped YLF and silica show potential
promise for thermal management, their low refractive index gives small
reflectivity, and thus small acceleration force. So, for lightsail
applications, we need to combine the laser-cooling layer with other
high-index materials~\cite{ilic2018nanophotonic}. As a representative
example, we consider a trilayer structure depicted in the inset of
\figref{power2}, with the aim to achieve sufficient reflection as well
as substantial cooling. The thickness of the two SiN layers at the
edges is chosen as small as possible while still maximizing reflection
for wavelengths around $1~\mu\textrm{m}$. The middle layer is either
silica or doped YLF and is used for thermal management. In all
computations, we assume a configuration of a typical sail area of
$10$~m$^2$ and a payload mass of $1$~kg. This configuration, where the
mass of the micron-thick sail is relatively small compared to the
kg-level payload, is adequate to reach a target velocity of few
percents the speed of light with a laser intensity of the order of few
$\textrm{GW/m}^2$ which corresponds to a Starshot-level
beam~\cite{lubin2016roadmap, parkin2018breakthrough}. In order to
quantify this, we compute the maximum lasing power
$I_{p,\mathrm{max}}$ and the corresponding acceleration distance $D$
required to reach a target velocity of $0.01 c$
(\figref{power2}). Here we do not assume a particular form of the SiN
absorption coefficient $\alpha_{\mathrm{SiN}}$ as it highly depends on
many processing-dependent
factors~\cite{ilic2018nanophotonic}. Instead, we assess the
performance of the lightsail at various values of
$\alpha_{\mathrm{SiN}}$. We also assume a fixed small absorption for
silica~\cite{ilic2018nanophotonic}
($\alpha_\textrm{silica}=10^{-6}~\textrm{cm}^{-1}$) and
Yb$^{3+}$-doped YLF ($\alpha_b=10^{-4}~\textrm{cm}^{-1}$ for the
state-of-art~\cite{melgaard2016solid} and
$\alpha_b=\alpha_\textrm{silica}$ for the ideal case) in the middle
layer. The specific choice of a threshold temperature $T_s$ depends on
the required operating conditions and the temperature dependence of
the material properties~\cite{atwater2018materials}. As a reference,
we now assume $T_s=300$~K (and later explore the effects of changing
$T_s$). The corresponding maximum allowed laser power
$I_{p,\mathrm{max}}$ is then computed by \eqref{balance}. For each
configuration, we fine-tune the operating wavelength around
$1\mu\textrm{m}$ to minimize the acceleration distance $D$.

\begin{figure*}[htbp]
  \centering
  \includegraphics[width=1\textwidth]{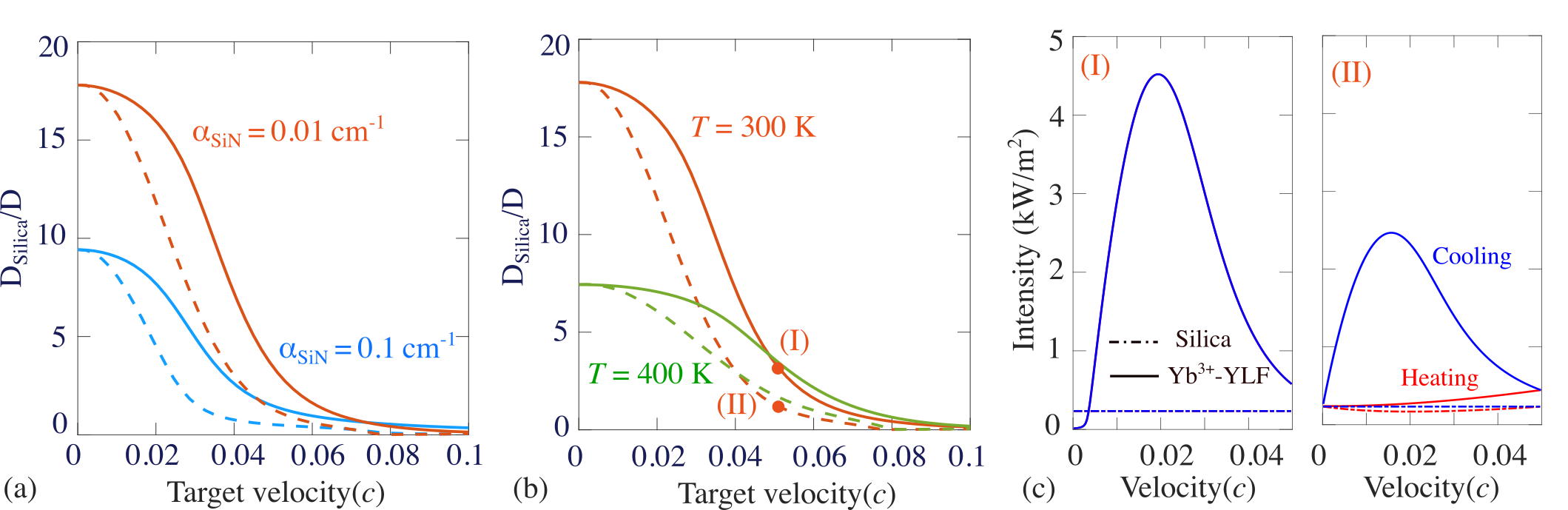}
  \caption{Results for the triple-layer setup shown in
    \figref{power2}(a) using a Yb$^{3+}$ doped YLF layer with small
    parasitic absorption ($\alpha_b=10^{-6}~\textrm{cm}^{-1}$) and
    ideal quantum yield ($\eta_e=100\%$). In (a,b), solid
    (resp. dashed) lines stand for tunable (resp. fixed) laser
    intensity during the acceleration stage.  (a) Improvement of
    acceleration distance as a function of target velocity for
    different SiN absorption coefficients and an operating temperature
    $T_s=300$~K. (b) Improvement of acceleration distance at different
    operating temperatures and a given SiN absorption
    $\alpha_{\mathrm{SiN}}=10^{-2}$~cm$^{-1}$. (c) Heating/cooling
    intensity during the acceleration process for target velocity
    $0.05c$ [red dots in (b)] corresponding to tunable (I) and fixed
    (II) laser intensity, using a laser-cooler Yb$^{3+}$-YLF layer
    (solid) or radiatively-cooled silica layer (dashed). Note that for
    tunable laser intensity (I), heating and cooling curves are
    identical.}
  \label{fig:velocity}
\end{figure*}

The results of the computation are shown in \figref{power2}. In
\figref{power2}(b), we see that substantial reduction of the
acceleration distance $D$ is possible when using laser cooling,
compared to using an undoped silica layer that only relies on thermal
radiation. As shown in \figref{power2}(a), this is due to the fact
that laser cooling allows the use of higher beam intensities without
exceeding the threshold temperature $T_s$. Note that, as expected,
this maximum intensity $I_{p,\mathrm{max}}$ always increases with
decreasing absorption coefficient $\alpha_{\mathrm{SiN}}$, and
saturates when the loss becomes dominated by the middle
layer. \figref{power2}(b) shows that, when using an ideal
Yb$^{3+}$-doped YLF layer (red solid curve), $D$ can be decreased by
large factors over a wide range of $\alpha_{\mathrm{SiN}}$ values,
reaching a maximum enhancement $D_{\textrm{silica}} / D \approx 22$
for small loss. For a state-of-art Yb$^{3+}$-YLF layer (red dashed
line), the larger parasitic absorption $\alpha_b$ of the middle layer
limits the enhancement at small $\alpha_{\mathrm{SiN}}$ values, since
optical absorption becomes quickly dominated by the middle
layer. Still, a large ($\approx 10 \times$) decrease of $D$ can be
observed for intermediate absorption coefficient values. Finally,
Yb$^{3+}$-doped silica (solid green line) gives smaller improvement of
at best a factor of $\approx 4$. The smaller improvement is expected
and is due to the corresponding smaller heat dissipation rate, as
already noted from the results of \figref{power}.

We further study the effects of changing the target velocity and the
threshold damage temperature $T_s$ when using a Yb$^{3+}$-YLF layer in
the optimal regime of laser cooling [\figref{velocity}(a,b)-dashed
lines]. As expected, the acceleration-distance enhancement
$D_{\textrm{silica}} / D$ worsens with increasing target velocity
since the operation bandwidth increases. In fact, because of Doppler
shift, the operating wavelength substantially shifts when the
lightsail reaches high velocities and can get outside the laser
cooling bandwidth. \figref{velocity}(a) [dashed lines] shows that $D$
increases and becomes larger than the acceleration distance
$D_{\mathrm{silica}}$ (corresponding to undoped silica) for target
velocities nearing 10\% of the speed of light. This effect is more
pronounced for larger absorption coefficients. Similarly,
\figref{velocity}(b) [dashed lines] reveals that laser cooling can
play a more significant role when aiming at a lower threshold
temperature due to the different temperature dependence for each heat
dissipation mechanism, where the laser cooling scales as $\propto T_s$
from \eqref{peak} and the thermal radiation as $\propto T_s^4$.

It is important to notice that, as the operating wavelength is Doppler
shifted during the acceleration phase, the laser cooling power changes
[e.g., \figref{scheme}(b)], meaning that the maximum laser intensity
allowed for the structure to stay at below the threshold temperature
$T_s$ also varies. For example, for velocities where the laser cooling
power is maximum, a larger laser intensity can be used without
exceeding the temperature $T_s$. If the operating laser intensity is
kept constant in the earth frame as the lightsail accelerates, then it
has to match the value of $I_{p,\textrm{max}}$ corresponding to the
lowest laser cooling power of the whole acceleration stage. So, in
order to estimate the benefits of a variable laser intensity, we
compute the corresponding enhancement in the acceleration distance
$D$, as plotted in \figref{velocity}(a,b) [solid lines]. We see that
noticeable improvement can be achieved, both as a function of the
target velocity and the threshold temperature. To highlight this, we
investigate the evolution of the cooling power during the acceleration
stage in multiple settings, for a target velocity of $0.05c$. When the
laser intensity is allowed to vary [\figref{velocity}(c,I)], heating
and cooling intensities are exactly balanced during the whole
acceleration stage. On the other hand, for a fixed laser intensity
[\figref{velocity}(c,II)], exact balance occurs only at specific
velocities (here, $0$ and $0.05c$). During most of the acceleration
stage, the lightsail sees a net cooling at the threhsold temperature
$T_s$, meaning that its equilibrium temperature is strictly less than
$T_s$. This allows for the laser intensity to be further increased
during most of the acceleration stage without exceeding the threshold
temperature $T_s$. It is this effect that is being exploited in a
variable laser intensity setting, which allows for an overall decrease
of the total acceleration distance $D$.

\section{Conclusion}

We have presented a general study for the prospect of using laser
cooling as a thermal management tool in relativistic lightsails. Our
results show the prospects of such method and the regime in which it
is most relevant (micron-thick structures, few percents the speed of
light, lower operating temperature, and appropriate absorption
interval). When aiming for even higher target velocity where the
limited operating bandwidth of laser cooling dopant to account for
Doppler shift becomes the main challenge, one can potentially still
harness laser cooling by using a laser of variable frequency, or using
a few rare earth ion to span a wide operating bandwidth. Additionally,
future work can focus on further design schemes to improve the
potential of using laser cooling in a lightsail setting.

\emph{Acknowledgments} W. Jin would like to thank Dr. Mohammed
Benzaouia and Dr. Alex Song for useful contributions. This work is
supported by the Breakthrough Starshot Initiative (initial idea and
performance testing), and the U.S. Department of Energy ``Photonics at
Thermodynamic Limits'' Energy Frontier Research Center under Grant
No. DE-SC0019140 (theoretical modeling).

\end{document}